\documentstyle[aps,twocolumn,epsfig]{revtex}

\begin{document}
\draft

\twocolumn[\hsize\textwidth\columnwidth\hsize\csname@twocolumnfalse\endcsname

\title{Thermal entanglement in three-qubit Heisenberg models}
\author{Xiaoguang Wang$^{1,2}$  Hongchen Fu$^{3}$  Allan I Solomon$^{3}$}
\address{1. Institute of Physics and Astronomy, University of Aarhus, DK-8000, Aarhus C, 
Denmark.}
\address{2. Institute for Scientific Interchange (ISI) Foundation, Viale Settimio Severo 65, 
I-10133 Torino, Italy}
\address{3. Quantum Processes Group, The Open University, Milton Keynes,  MK7 6AA, 
United Kingdom.}
\date{\today}
\maketitle

\begin{abstract}
We study  pairwise thermal entanglement 
in three-qubit  Heisenberg models and obtain 
analytic expressions for the concurrence.
We find  that  thermal entanglement is absent from both
 the antiferromagnetic
$XXZ$ model, and  the ferromagnetic $XXZ$ model with
anisotropy parameter $\Delta\ge 1$.
Conditions for the existence of thermal entanglement
are discussed in detail, as is the role of degeneracy and  the effects of
magnetic fields on  thermal entanglement 
and the quantum phase transition. 
Specifically, we find that the magnetic field can induce entanglement 
in the antiferromagnetic $XXX$ model,
but cannot induce entanglement in the ferromagnetic $XXX$ 
model.
\end{abstract}


]

\narrowtext

\section{Introduction}

\label{sec:intro}

Over the past few years much effort has been put into studying the
entanglement of multipartite systems both qualitatively and quantitatively.
Entangled states constitute  a valuable resource in quantum information
processing\cite{Bennett}. Quite recently, entanglement in quantum operations
\cite {EO0,EO1,EO2} and entanglement in indistinguishable fermionic and bosonic
systems\cite{EI1,EI2,EI3} have been considered. 
Entanglement in  two-qubit states
has been well studied in the literature, as have various kinds of
three-qubit entangled states\cite
{Dur,threeq,Rajagopal}. The three-qubit entangled states have been shown to possess
 advantages over the two-qubit states in  quantum teleportation\cite
{tele}, dense coding\cite{dense} and quantum cloning\cite{clone}. 

An interesting and natural type of entanglement, thermal
entanglement, was introduced and analysed within the Heisenberg $XXX$\cite{Arnesen},
$XX$\cite{Wang1}, and $XXZ$\cite{Wang2} models as well as the Ising model in a 
magnetic field\cite{Ising}. The state of the system at thermal equilibrium
is represented by the density operator 
$\rho (T)=\exp \left( -\frac H{kT}\right) /Z,$ where $Z=$tr$\left[ \exp
\left( -\frac H{kT}\right) \right]$ is the partition function, $H$ the
system Hamiltonian, $k$ is Boltzmann's constant which we henceforth take equal to 1, 
and $T$ the temperature. As 
$\rho (T)$ represents a thermal state, the entanglement in the state is
called {\em thermal entanglement}\cite{Arnesen}. A complication in the analysis is that, although  standard statistical
physics is characterized by the partition function, determined by the eigenvalues of the
system, thermal entanglement properties require in addition knowledge 
 of  the eigenstates.

The Heisenberg model  has been used to simulate a quantum computer\cite
{Loss}, as well as  quantum dots\cite{Loss}, nuclear spins\cite
{Kane}, electronic spins\cite{Vrijen} and optical lattices\cite{Moelmer}. By
suitable coding, the Heisenberg interaction alone can be used for quantum
computation\cite{Loss2}. The entanglement in
the ground state of the Heisenberg model has been discussed by O'Connor and
Wootters\cite{Oconnor}.

In  previous studies of thermal entanglement analytical results were
only available for two-qubit quantum spin models. 
In this paper  we   analyze
the three-qubit case, i.e. we consider  pairwise thermal entanglement in  three-qubit Heisenberg models. 

A general 3-qubit Heisenberg $XYZ$ model  in a non-uniform magnetic field {\bf $B$} is given by:
\begin{eqnarray}
H       &=&H_{XYZ}+H_{\rm mag} \nonumber \\
H_{XYZ} &=&\sum_{n=1}^3\left( \frac{J_1}{2}\sigma _{n}^{x}\sigma _{n+1}^{x}+\frac{J_2}{2}\sigma
_{n}^{y}\sigma _{n+1}^{y}+\frac{J_3}{2}\sigma
_{n}^{z}\sigma _{n+1}^{z}\right) \nonumber \\
H_{\rm mag} &=&\sum_{n=1}^3 {B_n \sigma _n^z}. 
\label{eq:xyz}
\end{eqnarray}
We use the standard notation, detailed later, and  assume a periodic boundary, identifying the subscript $4$ with $1$ in the above expressions. 
For the 3-qubit case even this most general scenario is susceptible to numerical analysis.  However, in this paper we shall restrict ourselves to special cases of Eq.(\ref{eq:xyz}) for which we are able to provide a succinct analytic treatment.

The 3-site Heisenberg models we will study in this paper are the following:
\begin{enumerate}
\item The $XX$ model, corresponding to $J_1=J_2,\;\;\; J_3=0$ and ${\bf B}=0$.
\item The $XXZ$ model, for which $J_1=J_2, J_3\neq 0$ and ${\bf B}=0$.
\item The $XXZ$ model with uniform magnetic field ($B_1=B_2=B_3$).
\end{enumerate}
We start in Sec. II by examining  the three-qubit $XX$ model. 
In Sec. III, IV, and V, we study thermal entanglement in the $XX$ model, the $XXZ$ model and the $XXZ$ model in a magnetic field, respectively.

During the course of the analysis it will become clear that  degeneracy plays an important
role in  thermal entanglement, as does the presence of magnetic fields. We find the  critical temperatures involved in the quantum phase transition associated with the existence of entanglement  in these  quantum spin models.

\section{Three-qubit $XX$ model and its solution}

The three-qubit $XX$ model is described by the Hamiltonian\cite{Lieb} 
\begin{eqnarray}
H_{XX} &=&\frac J2\sum_{n=1}^3\left( \sigma _{n}^{x}\sigma _{n+1}^{x}+\sigma
_{n}^{y}\sigma _{n+1}^{y}\right)   \nonumber \\
&=&J\sum_{n=1}^3\left( \sigma _{n}^{+}\sigma _{n+1}^{-}+\sigma _{n}^{-}\sigma
_{n+1}^{+}\right) ,  \label{xy1}
\end{eqnarray}
where $\sigma _n^\alpha $ $(\alpha =x,y,z)$ are the Pauli matrices of the $n$%
-th qubit, $\sigma _n^{\pm }=\frac 12\left( \sigma _n^x\pm i\sigma
_n^y\right) $ the raising and lowering operators, and $J$ is the exchange
interaction constant. Positive (negative) $J$ corresponds to the 
antiferromagnetic (ferromagnetic) case. As signalled  above, we adopt  periodic boundary conditions; 
$\sigma_4^x=\sigma_1^x,$ $\sigma_4^y=\sigma_1^y.$ 
We are therefore considering a three-qubit
Heisenberg ring. The $XX$ model was intensively investigated in
1960 by Lieb, Schultz, and Mattis\cite{Lieb}. More recently the $XX$ model has been 
realized in the quantum-Hall system\cite{Hall}, the cavity QED system\cite
{Zheng} and quantum dot spins\cite{I} for a quantum computer.

In order to  study thermal entanglement, the first step is to obtain all
the eigenvalues and eigenstates of  the Hamiltonian Eq.(\ref{xy1}). The eigenvalues
themselves do not suffice  to calculate the entanglement. The eigenvalue
problem of the $XX$ model can be exactly solved by the Jordan-Wigner
transformation\cite{JW}. In the three-qubit case the eigenvalues are more simply obtained as
\cite{Wang1}

\begin{eqnarray}
E_0 &=&E_7=0,  \nonumber \\
E_1 &=&E_2=E_4=E_5=-J,  \nonumber \\
E_3 &=&E_6=2J.  \label{eq:eeeigen}
\end{eqnarray}
\newline
and the corresponding eigenstates are explicitly given by

\begin{eqnarray}
|\psi _0\rangle &=&|000\rangle ,  \nonumber \\
|\psi _1\rangle &=&3^{-1/2}\left( q|001\rangle +q^2|010\rangle +|100\rangle
\right) ,  \nonumber \\
|\psi _2\rangle &=&3^{-1/2}\left( q^2|001\rangle +q|010\rangle +|100\rangle
\right) ,  \nonumber \\
|\psi _3\rangle &=&3^{-1/2}\left( |001\rangle +|010\rangle +|100\rangle
\right) ,  \nonumber \\
|\psi _4\rangle &=&3^{-1/2}\left( q|110\rangle +q^2|101\rangle +|011\rangle
\right) ,  \nonumber \\
|\psi _5\rangle &=&3^{-1/2}\left( q^2|110\rangle +q|101\rangle +|011\rangle
\right) ,  \nonumber \\
|\psi _6\rangle &=&3^{-1/2}\left( |110\rangle +|101\rangle +|011\rangle
\right) ,  \nonumber \\
|\psi _7\rangle &=&|111\rangle .  \label{eq:estate}
\end{eqnarray}
with $q=\exp \left( i2\pi /3\right) $ satisfying

\begin{eqnarray}
q^3 &=&1,  \nonumber \\
q^2+q+1 &=&0.
\end{eqnarray}
This set (\ref{eq:estate}) of three-qubit states is itself interesting.
Rajagopal and Rendell\cite{Rajagopal} have considered a similar set of  three-qubit states which they have  classified by means of 
permutation symmetries. Here the states $|\psi _0\rangle ,|\psi _3\rangle ,$ 
$|\psi _6\rangle ,$ and $|\psi _7\rangle $ are symmetric in the permutation
of any pair of particles. We define a cyclic shift operator $P$ by its
action on the basis $|ijk\rangle $\cite{Schnack}

\begin{equation}
P|ijk\rangle =|kij\rangle .
\end{equation}
Obviously the four states $|\psi _0\rangle ,|\psi _3\rangle ,$ $|\psi
_6\rangle ,$ and $|\psi _7\rangle $ are the eigenstates of $P$ with
eigenvalue 1. The other four states in the set (\ref{eq:estate}) are also
eigenstates of $P$ as follows:

\begin{mathletters}
\begin{eqnarray}
P|\psi _i\rangle &=&q^2|\psi _i\rangle \text{ }(i=1,4), \\
P|\psi _j\rangle &=&q|\psi _j\rangle \text{ }(j=2,5).
\end{eqnarray}
\end{mathletters}
This is not surprising since the Hamitonian $H_{XX}$ as well as the other 
Hamiltonians considered later are  invariant under 
the cyclic shift operator.

For $J>0$ ($J<0$) the ground state is four (two)-fold degenerate. We will
see that the degeneracy of the system  influences thermal entanglement greatly. There is no pairwise entanglement in the eigenstate $|\psi
_0\rangle $ and $|\psi _7\rangle .\,$\thinspace Pairwise entanglement
exists in the state $|\psi _i\rangle $ ($i=1,2,...,6$) and the concurrence 
between any two different qubits is given by 2/3\cite{Dur,Koashi}.

\section{Thermal entanglement in the $XX$ model}

We first recall the definition of {\em concurrence}\cite{Wootters1} between a pair of qubits. Let $\rho _{12}$ be the density matrix of the pair and it can be either pure or mixed. The concurrence corresponding to the density matrix is
defined as 
\begin{equation}
{\cal C}_{12}=\max \left\{ \lambda _1-\lambda _2-\lambda _3-\lambda
_4,0\right\} ,  \label{eq:c1}
\end{equation}
where the quantities $\lambda _i$ are the square roots of the eigenvalues of
the operator 
\begin{equation}
\varrho _{12}=\rho _{12}(\sigma _1^y\otimes \sigma _2^y)\rho _{12}^{*}(\sigma
_1^y\otimes \sigma_2^y)  \label{eq:c2}
\end{equation}
in descending order. The eigenvalues of $\varrho _{12}$ are 
real and non-negative even though $\varrho _{12}$ is not necessarily
Hermitian. The values of the concurrence range from zero, for an unentangled
state, to one, for a maximally entangled state.

The state at thermal equilibrium is described by the density matrix
\begin{eqnarray}
\rho (T) &=&\frac 1Z\exp \left( -\beta H\right) ,  \nonumber \\
&=&\frac 1Z\sum_{k=0}^7\exp \left( -\beta E_k\right) |\psi _k\rangle \langle
\psi _k|.  \label{eq:rhoo}
\end{eqnarray}
where $\beta =1/T.$ From Eq.(\ref{eq:eeeigen}), the partition function is
obtained as

\begin{equation}
Z=2+4e^{\beta J}+2e^{-2\beta J}.  \label{eq:z}
\end{equation}
>From Eqs.(\ref{eq:eeeigen}) and (\ref{eq:rhoo}), the density matrix can be
written as

\begin{eqnarray}
\rho (T) &=&\frac 1Z[|\psi _0\rangle \langle \psi _0|+|\psi _7\rangle
\langle \psi _7|  \nonumber \\
&&+e^{\beta J}(|\psi _1\rangle \langle \psi _1|+|\psi _4\rangle \langle \psi
_4|  \nonumber \\
&&+|\psi _2\rangle \langle \psi _2|+|\psi _5\rangle \langle \psi _5|) 
\nonumber \\
&&+e^{-2\beta J}(|\psi _3\rangle \langle \psi _3|+|\psi _6\rangle \langle
\psi _6|)].  \label{eq:re}
\end{eqnarray}

In this paper we consider only pairwise thermal entanglement, and so we need to
calculate the reduced density matrix $\rho_{12}(T)=\text{tr}_3(\rho(T))$. 
We denote the reduced density matrix tr$_3[|\psi _{i_1}\rangle \langle
\psi _{i_1}|+...+|\psi _{i_N}\rangle \langle \psi _{i_N}|]$ by  $\rho
_{12}^{(i_1i_2...i_N)}.$   From Eq.(\ref{eq:estate}), we obtain

\begin{mathletters}
\begin{eqnarray}
\rho _{12}^{(07)} &=&\left( 
\begin{array}{llll}
1 & 0 & 0 & 0 \\ 
0 & 0 & 0 & 0 \\ 
0 & 0 & 0 & 0 \\ 
0 & 0 & 0 & 1
\end{array}
\right) , \label{eq:aaa}\\
\rho _{12}^{(1245)} &=&\frac 23\left( 
\begin{array}{llll}
1 & 0 & 0 & 0 \\ 
0 & 2 & -1 & 0 \\ 
0 & -1 & 2 & 0 \\ 
0 & 0 & 0 & 1
\end{array}
\right) , \label{eq:bbb}\\
\rho _{12}^{(36)} &=&\frac 23\left( 
\begin{array}{llll}
\frac 12  & 0 & 0 & 0 \\ 
0 & 1 & 1 & 0 \\ 
0 & 1 & 1 & 0 \\ 
0 & 0 & 0 & \frac 12 
\end{array}
\right),\label{eq:ccczzz}\\
\rho _{12}^{(012)} &=&\frac 23\left( 
\begin{array}{llll}
\frac 52 & 0 & 0 & 0 \\ 
0 & 1 & -\frac 12  & 0 \\ 
0 & -\frac 12  & 1 & 0 \\ 
0 & 0 & 0 & 0 
\end{array}
\right) ,\label{eq:ddd}\\
\rho _{12}^{(12)} &=&\frac 23\left( 
\begin{array}{llll}
1 & 0 & 0 & 0 \\ 
0 & 1 & -\frac 12  & 0 \\ 
0 & -\frac 12  & 1 & 0 \\ 
0 & 0 & 0 & 0
\end{array}
\right).\label{eq:eee} 
\end{eqnarray}
The last two reduced density matrices will be used  
later.

>From Eqs. (\ref{eq:re}) and (\ref{eq:aaa}-\ref{eq:ccczzz}), we obtain 
\end{mathletters}
\begin{eqnarray}
\rho _{12}(T) &=&\frac 1Z\left( \rho _{12}^{(07)}+e^{\beta J}\rho
_{12}^{(1245)}+e^{-2\beta J}\rho _{12}^{(36)}\right)   \nonumber \\
&=&\frac 2{3Z}\left( 
\begin{array}{llll}
v & 0 & 0 & 0 \\ 
0 & w & y & 0 \\ 
0 & y & w & 0 \\ 
0 & 0 & 0 & v
\end{array}
\right)   \label{eq:vwy}
\end{eqnarray}
with
\begin{eqnarray}
v &=&\frac 32+e^{\beta J}+\frac 12e^{-2\beta J},  \nonumber \\
w &=&2e^{\beta J}+e^{-2\beta J},  \nonumber \\
y &=&e^{-2\beta J}-e^{\beta J}.
\end{eqnarray}

The square roots of the four eigenvalues of the operator $\varrho _{12}$ are
\begin{eqnarray}
\lambda _1 &=&\frac{2(w-y)}{3Z},\text{ }\lambda _2=\frac{2(w+y)}{3Z}, 
\nonumber \\
\lambda _3 &=&\lambda _4=\frac{2v}{3Z}.  \label{eq:lam}
\end{eqnarray}
>From Eqs.(\ref{eq:c1}), (\ref{eq:z}), (\ref{eq:vwy}), and (\ref{eq:lam}), we
obtain the concurrence\cite{Oconnor}

\begin{eqnarray}
{\cal C}&=&\frac 4{3Z}\max (|y|-v,0),  \label{eq:ccc} \\
&=&\max \left[ \frac{2|e^{-2x}-e^x|-3-2e^x-e^{-2x}}{%
3(1+2e^x+e^{-2x})},0\right]   \label{eq:cc1}
\end{eqnarray}
where $x \equiv \beta J = J/T$. The concurrence 
depends only on the {\em ratio} of $J$ and $T$.
Due to symmetry under cyclic shifts, the value of the concurrence does not depend on the choice of the  pair of qubits.

>From (\ref{eq:cc1}) we see that entanglement
appears only when
\begin{equation} 
\frac{2|z^{-2}-z|-3-2z-z^{-2}}{%
3(1+2z+z^{-2})} >0,
\end{equation}
or in other words
\begin{equation} \label{fenmu}
2|z^{-2}-z|- 3 - 2 z- z^{-2} >0,
\end{equation}
where $z=\exp(x)$.
We now consider two different cases:

{\it Case 1}. Antiferromagnetic system; $J>0$;$z^{-2}-z<0$. In this case relation (\ref{fenmu}) requires
\begin{equation}
z^{-2}<-1
\end{equation}
which is impossible. So there is no entanglement when
$J>0$.

{\it Case 2}. Ferromagnetic system; $J<0$,$z^{-2}-z > 0$. Relation (\ref{fenmu}) becomes
\begin{equation}
z^{-2}-4 z -3 >0.
\end{equation}
or
\begin{equation} 
f(z) \equiv 4z^3+3z^2-1<0.  \label{yfunction}
\end{equation}
The function $f(z)$ is an increasing function of the positive
real argument $z$ and relation (\ref{yfunction}) is valid iff 
$0<z<z_c$, where the critical value $z_c$  determined by $f(z_c)=0$
is 0.4554; that is, $ x<-0.7866 $. For fixed $J$, we obtain the
critical temperature $T_c=1.21736|J|$, above which  there is no thermal entanglement. The critical temperature  depends linearly  on the absolute value of $J.$

In the ferromagnetic case the concurrence 
\begin{equation}
{\cal C}= \max\left[\frac{1-4z^3-3z^2}{3(1+2z^3+z^2)},0\right] \label{eq:cc111}
\end{equation}
reaches its maximum value of  $1/3$ when $z \to 0$, that is when 
$x\to -\infty$. Since the entanglement is a monotonic increasing function of ${\cal C}$ this means that the entanglement attains its maximum value for zero temperature, when  $J$  is finite and nonzero. 
For finite temperatures, this maximum is also attained when $J\to -\infty$. 

In summary, we find that \\ \\ 
{\bf Theorem 1:} {\it The XX model is thermally  entangled if and 
only if  $J<-0.7866 T$; maximum entanglement is attained when $T \to 0$ or $J\to -\infty$.}

The above discussion shows  that  in our 3-qubit model pairwise thermal entanglement occurs only in the ferromagnetic case. This result  differs from that for the two-qubit $XX$ model, for  which thermal entanglement exists in both the antiferromagnetic and ferromagnetic cases\cite{Wang1}.

For the ferromagnetic case 
the states $|\psi _3\rangle $ and $|\psi_6\rangle $ constitute a  doubly-degenerate ground state.
Eq.(\ref{eq:cc111}) shows that the concurrence ${\cal C}=1/3$ 
at zero temperature. As noted in the last section the concurrence for any
two qubits in the state $|\psi _3\rangle $ or $|\psi _6\rangle $ is 2/3.
Here the value 1/3 appears  due to the degeneracy. In fact, at zero temperature,
the thermal entanglement can be calculated from $\rho^{(36)}_{12}$(\ref{eq:eee}). 
After normalization it is easy to
obtain the concurrence, which is just 1/3.

\section{The Anisotropic Heisenberg $XXZ$ model}
We now consider a more general Heisenberg model, the anisotropic Heisenberg $XXZ$ model,  which is
described by the Hamiltonian\cite{A} 
\begin{equation}
H_{XXZ}=H_{XX}+\frac{\Delta J}2\sum_{n=1}^3(\sigma _n^z\sigma _{n+1}^z-1),
\end{equation}
where $\Delta $ is the anisotropy parameter. The model reduces to the $XX$
model when $\Delta =0,$ and the isotropic Heisenberg $XXX$ model when $\Delta
=1.$

It is straightforward to check that the added anisotropic term $H_{XXZ}-H_{XX}$
commutes with $H_{XX}$. Therefore the eigenstates of the $XXZ$ model
are still given by Eq. (\ref{eq:estate}), now with the different
eigenvalues 
\begin{eqnarray}
E_0 &=&E_7=0,  \nonumber \\
E_1 &=&E_2=E_4=E_5=-2J(\Delta +\frac 12),  \nonumber \\
E_3 &=&E_6=-2J(\Delta -1).
\end{eqnarray}

Following the procedure of the previous section, 
we obtain the concurrence, which is
of the same form as Eq.(\ref{eq:ccc}) with however the parameters $v,w,y,$
and the partition function $Z$ now given by
\begin{eqnarray}
v &=&\frac 32+\frac 12z^{2\Delta }(2z+z^{-2}), 
\nonumber \\
w &=&z^{2\Delta }(2z+z^{-2}),  \nonumber \\
y &=&z^{2\Delta }(z^{-2}-z),  \nonumber \\
Z &=&2+2z^{2\Delta }(2z+z^{-2}).  \label{eq:vvv}
\end{eqnarray}

As in the last section, since $Z$ is always positive, we need only 
 consider
\begin{eqnarray}
f(\Delta,z) 
&\equiv& |y|-v \nonumber \\
&=& z^{2\Delta}|z^{-2}-z|-\frac{3}{2}-
z^{2\Delta+1}-\frac{1}{2}z^{2\Delta-2} 
\end{eqnarray}
to determine whether  entanglement occurs or not.
Again, we have to consider two
different cases:

{\em Case 1}. When $J>0$ ($z>1$), namely the 
antiferromagnetic $XXZ$ model, the condition on $f(\Delta,z)$ leads to 
\begin{equation}
z^{2\Delta-2}=e^{2x(\Delta-1)}<-1,
\end{equation}
which is impossible. So there is no entanglement in this 
case, irrespective of $\Delta$.

{\em Case 2}. When $J<0$ ($z<1$), namely the 
ferromagnetic $XXZ$ model, the condition $f(\Delta,z)>0$ gives
\begin{equation} \label{Jsmaller0}
z^{2\Delta-2}-4 z^{2\Delta+1} -3 >0.
\end{equation}
We consider some special values of $\Delta$.

(1) $\Delta\ge 1$: For $\Delta=1$ the 
relation (\ref{Jsmaller0})
implies $z^3<-1/2$ which is impossible. So there is
no entanglement in the  $XXX$ model. We can 
further prove that there is no entanglement for 
$\Delta>1$. In fact, it is easy to see that
\begin{equation}
f(\Delta,z)<z^{2(\Delta-1)}-3<0,
\end{equation}
where we have used the inequalities  $z^{2\Delta+1}>0$
and $z^{2(\Delta-1)}<1$ for $\Delta>1$ and
$z<1$. This means ${\cal C}=0$ and thus there is no 
entanglement.

(2) $\Delta=1/2$: In this
case\cite{Plus} the entanglement condition is obtained as
\begin{equation}
4z^3+3z-1<0,
\end{equation}
which is an increasing function of $z$. So the
model is entangled iff $0<z<z_c\approx 0.298$, where 
$z_c$ is determined as a root of $4z^3+3z-1=0$.

(3) $\Delta=-1/2$: This is an interesting case whose importance has been emphasized recently\cite{Minus}.
>From the eigenvalues we see that the excited state of the system
is 6-fold degenerate when $\Delta=-1/2$.
The function $f(\Delta,z)$ now reduces to $z^{-3}-7$, from which
the critical values $z_c$ and $T_c$ are obtained analytically
as $z_c=7^{-1/3}, T_c={3}|J|/{\ln 7}\approx 1.5417|J|$.

(4) The limit case $\Delta\to -\infty$: The critical
value $z_c$ is now determined by $z^{-2}-4z=0$, i.e., $z_c=4^{-1/3}$. 
Therefore the critical temperature $T_c=3|J|/\ln 4\approx 2.164 |J|$.

Finally, for more general values of the anisotropy parameter we need to resort to  numerical calculations. Fig.1 is a plot of the critical temperature as a function of the anisotropy parameter $\Delta$. From this we see that the critical temperature decreases as $\Delta$ increases, and reaches the asymptotic value $T_c=2.1640|J|$
as $\Delta \rightarrow -\infty .$

\begin{figure}[tbh]
\begin{center}
\epsfxsize=8cm
\epsffile{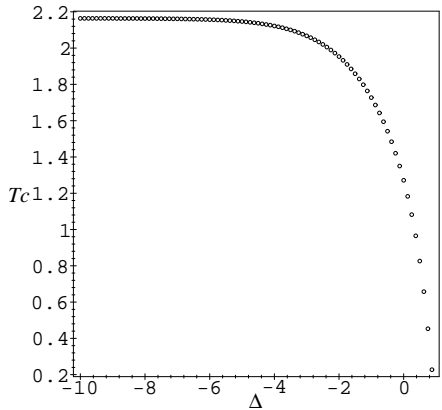}
\end{center}
\caption{The critical temperature $T_c$ as a function of the anisotropy
parameter $\Delta$. The exchange  constant $J=-1$.}
\end{figure}

We now give further  analytical results for the 
case $\Delta<1$ and $z<1$. 
Consider $f(\Delta, z)$ as a function of $\Delta$. Then,
from
\begin{eqnarray}
\frac{\partial f(\Delta, z)}{\partial \Delta}
&=&(\ln z) z^{2\Delta} (z^{-2}-4z) \nonumber\\
&& \left\{\begin{array}{ll}
=0, & \mbox{when } z=z_0\equiv 4^{-1/3}\approx 0.62996; \\
>0, & \mbox{when } z>z_0; \\
<0, & \mbox{when } z<z_0.
\end{array} \right.
\label{eq:gradients} \end{eqnarray}
we see that $f(\Delta, z)$
is an increasing (decreasing) function when $z>z_0$
($z<z_0$). We consider these cases separately.

{\em Case 2a}. When $z=z_0$, $f(\Delta, z_0)=-3<0$.
So there is no entanglement in this case.

{\em Case 2b}. When $z>z_0$, the function $f(\Delta, z)$
is an increasing function which reaches its maximum when
$\Delta\to 1$. Since we have seen that  there is no entanglement
when $\Delta=1$
\begin{equation}
f(\Delta, z)<f(1, z)<0 \quad \mbox{for }z>z_0,
\end{equation}
which means that there is no entanglement when $z>z_0$.

{\em Case 2c}. The case $z<z_0$. 
Define the $z$-dependent  point $\Delta_z$ by  $f(\Delta_z, z)=0$
where
\begin{equation}
\Delta_z=\frac{1}{2\beta J} \ln \left[
\frac{3}{z^{-2}-4z}\right]<1.
\label{eq:delta} \end{equation} 
Thus from Eq.(\ref{eq:gradients}) we know that $f(\Delta, z)>0$ when $\Delta<\Delta_z$ for
all $z<z_0$, which is just the condition for 
entanglement.

 In Fig.\,2 we give plots of $f(\Delta, z)$
for $z=0.6295, 0.6, 0.5, 0.4, 0.3, 0.2, 0.1$. Note that 
$\Delta_z$ is a decreasing function of $z$ and that
\begin{equation}
\Delta_z \to \left\{
\begin{array}{ll}
-\infty & \mbox{when } z\to z_0; \\
1       & \mbox{when } z\to 0,
\end{array} \right.
\end{equation}
as indicated in Fig.\,2.

\begin{figure}[tbh]
\begin{center}
\epsfxsize=8cm
\epsffile{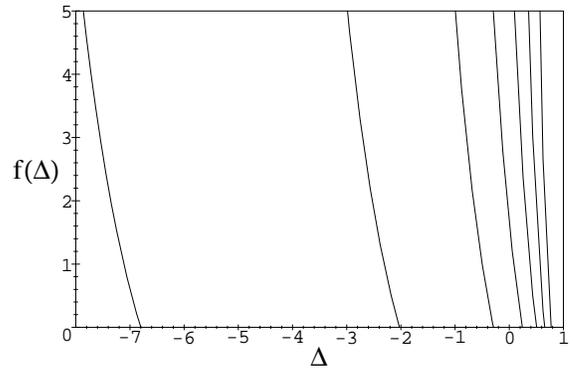}
\end{center}
\caption{The function $f(\Delta, z)$ with respect $\Delta$ for $z=0.6295, 
0.6, 0.5, 0.4, 0.3, 0.2, 0.1$ (from left to right).}
\end{figure}

In summary, we have \\ \\
{\bf Theorem 2}
{\it The XXZ model exhibits thermal entanglement only when
$z<z_0$, that is, $J<-.4621T$,  and $\Delta<\Delta_z$.}

Note that Theorem 2 is entirely consistent with Theorem 1, since $\Delta_z=0$ in Eq.(\ref{eq:delta}) corresponds to $J=-0.7866T.$

In Fig.3 we plot 
the concurrence as a function of the anisotropy parameter 
$\Delta $ and exchange constant $J$. The figure shows that there is  no thermal
entanglement for the antiferromagnetic ($J>0$) $XXZ$ model, nor for 
the ferromagnetic ($J<0$) $XXZ$ model when $\Delta \ge 1$. 

\begin{figure}[tbh]
\begin{center}
\epsfxsize=8cm
\epsffile{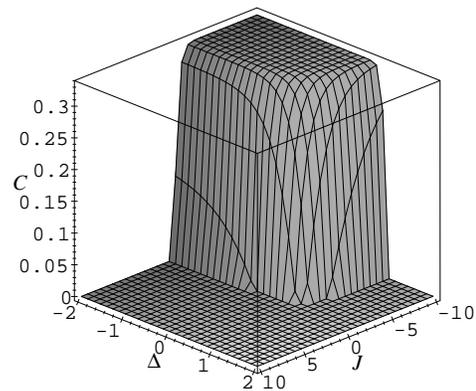}
\end{center}
\caption{The concurrence as a function of $\Delta$ and $J$. The temperature $T=1$.}
\end{figure}

To end  this section we investigate the  concurrence at 
zero temperature. Nonzero concurrence occurs for the case $\Delta<1, J<0$. 
In this case, the doubly-degenerate ground state consists of 
$|\psi_3\rangle $ and $|\psi_6\rangle $. We may calculate  the concurrence  ${\cal C} = 1/3$ directly from the density matrix
$\rho^{(36)}_{12}$.

\section{Effects of magnetic fields}
In this section we consider the effect of magnetic fields on thermal
entanglement. The $XXZ$ model with uniform magnetic field $B$ along the $z$ direction
is given by
\begin{equation}
H_{XXZM}=H_{XXZ}+B\sum_{n=1}^3\sigma _n^z.
\end{equation}
It is easy to check that the added magnetic term commutes with the Hamiltonian
$H_{XXZ}$. Therefore the eigenstates of the $XXZ$ model
are given by Eq.(\ref{eq:estate}). The eigenvalues are now 
\begin{eqnarray}
E_0 &=&-3B  \nonumber \\
E_1 &=&E_2=-2J(\Delta +\frac 12)-B  \nonumber \\
E_3 &=&-2J(\Delta -1)-B,  \nonumber \\
E_4 &=&E_5=-2J(\Delta +\frac 12)+B,  \nonumber \\
E_6 &=&-2J(\Delta -1)+B.  \nonumber \\
E_7 &=&3B\label{eq:eigen}
\end{eqnarray}
We see that the magnetic field partly removes the degeneracy. 

With a  derivation completely analogous to that of  Sec. III and Sec. IV, the reduced
density operator is 
\begin{equation}
\rho _{12}=\frac 2{3Z}\left( 
\begin{array}{llll}
u & 0 & 0 & 0 \\ 
0 & w & y & 0 \\ 
0 & y & w & 0 \\ 
0 & 0 & 0 & v
\end{array}
\right)
\end{equation}
with

\begin{eqnarray}
u &=&\frac 32e^{3\beta B}+\frac 12 e^{\beta B}z^{2\Delta}\left(
2z+z^{-2}\right) ,  \nonumber \\
v &=&\frac 32e^{-3\beta B}+\frac 12e^{-\beta B}z^{2\Delta}\left(
2z+z^{-2}\right) ,  \nonumber \\
w &=&\cosh (\beta B)z^{2\Delta }\left( 2z+z^{-2}\right) ,  \nonumber \\
y &=&\cosh (\beta B)z^{2\Delta }\left(z^{-2}-z\right) , \nonumber \\
Z &=&2\cosh (3\beta B)+2\cosh (\beta B)z^{2\Delta}  \nonumber \\
&&\times (2z+z^{-2})
\end{eqnarray}

The concurrence is then given by

\begin{equation}
{\cal C}=\frac 4{3Z}\max (|y|-\sqrt{uv},0).
\end{equation}
As an immediate consequence we see that  the concurrence is an even function
of the magnetic field.

As the quantities $Z, u, v$ are all positive, for
convenience we consider the quantity $y^2-uv$ instead of $|y|-\sqrt{uv}$. Thermal entanglement occurs when
\begin{equation}
y^2-uv=h\cosh (2\beta B)-g>0,
\end{equation}
where
\begin{eqnarray}
&&g=\frac{1}{4}[9+z^{4(\Delta -1)}(2z^6+8z^3-1)],  \nonumber \\
&&h=\frac{1}{2}z^{2\Delta }\left[ z^{2\Delta }(z^{-2}-z)^2-(6z+3z^{-2})\right]
.  \label{eq:yyuv}
\end{eqnarray}

We now consider the effect of a magnetic field on  the thermal 
entanglement.

We first consider the $XXX$ model, $\Delta =1$, which
does not exhibit thermal entanglement when $B=0$. 
One might  expect that the magnetic field would induce 
thermal entanglement. It is easy to see that 
\begin{equation}
2(y^2-uv)=\cosh (2\beta B)\left( z^6-8z^3-2\right) -(z^3+2)^2.
\end{equation}
If $z<(4+3\sqrt{2})^{1/3}\approx 2.02$, $z^6-8z^3-2<0$ and thus 
$y^2-uv<0$ for any $B$. So for this range of $z$ values there is no thermal
entanglement no matter how strong the magnetic field is. 
However, when $z>(4+3\sqrt{2})^{1/3}$, 
$z^6-8z^3-2>0$ and the condition for entanglement 
becomes
\begin{equation}
\cosh (2\beta B)>\frac{(z^3+2)^2}{z^6-8z^3-2}
\end{equation}
which can be fulfilled for strong enough $B$. So a magnetic
field can induce  entanglement in the $XXX$ model  when
$z>(4+3\sqrt{2})^{1/3}$.

Now  consider the case $\Delta =-1/2$. 
>From Eq.(\ref{eq:yyuv}) we obtain
\begin{eqnarray}
h &=&\frac 12(p^2-5p-5), \\
g &=&\frac 14(11+8p-p^2), \\
h-g &=&\frac 14(3p^2-18p-21),
\end{eqnarray}
which are  parabolas in $p\equiv z^{-3}$,
as shown in Fig.\,4. We consider three different cases:

{\em Case 1: } $p<p_1=5/2+3\sqrt{5}/2$, 
In this case $h<0,g>0,h-g<0$ and $y^2-uv<0$. So there is
no thermal entanglement.

{\em Case 2:} $p_1<p<p_2=7$. In this case 
$h>0,g>0,h-g<0$. So $y^2-uv>0$, and so
entanglement appears if the magnetic field
is strong enough.  

{\em Case 3:} $p_2<p$.  In this case $h>0, h-g>0$ and $y^2-uv$
is always positive; that is, here the XXZ model
exhibits thermal entanglement for any magnetic 
field. Note that $p_2 = z^{-3}_c$ where $z_c$ is
the critical value given in last section. 

The above two models show that the magnetic field can
either induce entanglement in a non-entangled system 
or extend the entanglement range for an already entangled system.


\begin{figure}[tbh]
\begin{center}
\epsfxsize=8cm
\epsffile{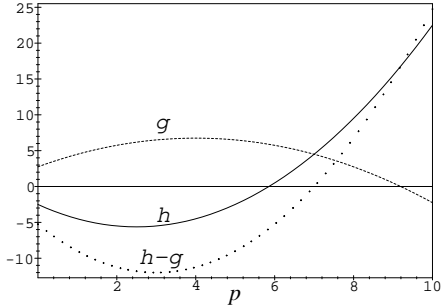}
\end{center}
\caption{The functions $h$, $g$, and $h-g$ in terms of $p=z^{-3}$.}
\end{figure}

In Fig.5 we plot the concurrence as a function of the magnetic field $B$ and exchange constant $J.$ 
At $B=0$ there is no thermal entanglement.
The entanglement increases with the magnetic field $|B|$ until it 
reaches a maximum value, then decreases and gradually disappears.
We can clearly see that there is no thermal entanglement for the
ferromagnetic case, while  thermal entanglement exists for the
antiferromagnetic case.
In other words,  we can induce entanglement in  the
antiferromagnetic $XXX$ system by introducing a  magnetic field, but cannot induce entanglement in the 
ferromagnetic $XXX$ system for any strength of magnetic field.

\begin{figure}
\begin{center}
\epsfxsize=8cm
\epsffile{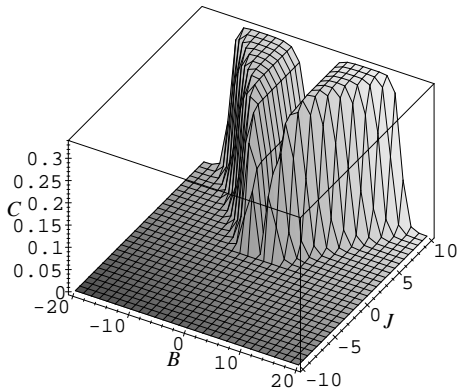}
\end{center}
\caption{
Concurrence as a function of the magnetic field $B$ and the exchange
constant $J$. The temperature $T=1$ and the anisotropy parameter $\Delta=1$.
}
\end{figure}

\begin{figure}[tbh]
\begin{center}
\epsfxsize=8cm
\epsffile{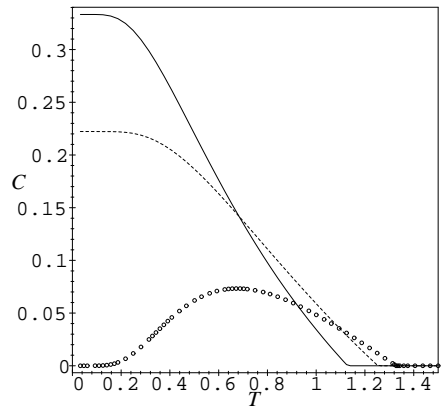}
\end{center}
\caption{
Concurrence as a function of $T$ for different magnetic fields $B=1$(solid line), 3/2(dashed line), and 2(circle point line).
}
\end{figure}

Fig. 6 gives a plot of the concurrence as a function of the temperature for
different magnetic fields. One can see that there exist critical
temperatures above which the entanglement vanishes. 
It is also noteworthy that the critical temperature increases 
as the magnetic field $B$ increases.
Consider the interesting case $B=2$.
We observe that the concurrence is zero at zero temperature and there is 
a maximum value of 
concurrence at a finite temperature. The entanglement can be increased by
increasing the temperature. The maximum value is due to the optimal mixing of all 
eigenstates in the system. When considering zero temperature we 
find that there are different limits for
different magnetic fields. Actually a more general result exists
\begin{eqnarray*}
\lim_{T\rightarrow 0}{\cal C}(\Delta ,B,1,T) &=&\frac 13\text{ for }\Delta
>|B|-1/2. \\
&=&\frac 29\text{ for }\Delta =|B|-1/2, \\
&=&0\text{ for }\Delta <|B|-1/2.
\end{eqnarray*}

The special point $T=0, \; \; \Delta=B-1/2$ 
($B\ge 0$ is assumed without loss of generality), at which the entanglement
undergoes a sudden change with adjustment of  the parameters $\Delta$ and $B$, is the
point of quantum phase transition\cite{QPT}. The quantum phase transition
takes place at zero temperature due to the variation of interaction terms
in the Hamiltonian. By examining the eigenvalues (\ref{eq:eigen}) we can understand the
phase transition. 
When $\Delta=B-1/2$, the ground state contains 
the three-fold degenerate states $|\psi_0\rangle, 
|\psi_1\rangle$, and $|\psi_2\rangle$. 
One may calculate the thermal entanglement 
from the density matrix $\rho_{12}^{(012)}$(\ref{eq:ddd}) and find the concurrence to be  $2/9$.
When $\Delta>B-1/2$, the ground state contains 
the two-fold degenerate states 
$|\psi_1\rangle$ and $|\psi_2\rangle$. The concurrence has the value $1/3$ as  calculated from 
$\rho_{12}^{(12)}$(\ref{eq:eee}). When $\Delta<B-1/2$, the ground state is $|\psi_0\rangle$ and
not degenerate. And the concurrence is zero in this case. 

\section{Conclusions}
Apart from being a fundamental property of quantum mechanics, it appears
that entanglement may provide an important resource in quantum information
processes. One source of entanglement is provided by magnetic systems, such
as those modelled in this paper. Within the current state of knowledge, only
measures for {\em pairwise} entanglement are available. Thus, in order to
study the entanglement properties of systems more complex than those simply
involving two qubits, it is necessary to adopt a procedure whereby one
traces out a subsystem, leaving effectively only a two-qubit system for
which we can calculate the {\em concurrence}, which in turn gives a  measure
of the entanglement.   Using this procedure, we have studied pairwise 
thermal entanglement in the following  Heisenberg models;
the  $XX$ model, the  $XXZ$ model and the $XXZ$ model in a  magnetic field.       
We obtained
analytical expressions for the concurrence,  which indicated
no thermal entanglement for the antiferromagnetic $XXZ$ model, nor for  the
ferromagnetic $XXZ$ model when the anisotropy parameter $\Delta \ge 1$. 
Conditions for the existence 
of thermal entanglement were studied in detail. 
The effects of magnetic fields on  entanglement were also considered. We found that 
the magnetic field can induce entanglement in the antiferromagnetic $XXX$ model,
but cannot induce entanglement in the ferromagnetic $XXX$ model, no matter how
strong the magnetic field is. 

In this paper we have extended previous work on  thermal entanglement
from two-qubit models to three qubit models, concentrating 
on those systems where the pairwise entanglement can be 
studied analytically. It would be an attractive 
proposition to extend further the investigation 
of such Heisenberg models to the  $N$-qubit case, which are under consideration.
\acknowledgments

X. Wang thanks K. M\o lmer, A. S\o rensen, W. K. Wootters and Paolo Zanardi for many
valuable discussions. He is supported by the Information Society
Technologies Programme IST-1999-11053, EQUIP, action line 6-2-1 and European project 
Q-ACTA. H.\,Fu is supported in part by 
the National Natural Science Foundation of China
(19875008), and A.\,I.\,Solomon 
acknowledges the hospitality of the 
Laboratoire de Physique Th\'{e}orique des Liquides, 
Paris University VI.

\end{document}